\documentclass{iopart}

\usepackage{iopams,setstack,amssymb,latexsym}
\usepackage{graphicx}
\usepackage{dcolumn}
\usepackage{bm}

\def \be{\begin{equation}}
\def \ee{\end{equation}}
\def \bea{\begin{eqnarray}}
\def \eea{\end{eqnarray}}
\def \nn{\nonumber}
\def \ie{{\it i.e.}}
\def \eg{{\it e.g.}}
\def \d{{\rm d}}
\def \del{\partial}
\def \tr{{\rm tr}}

\begin{document}

\title{Quantum Aspects of Massive Gravity}
\date{\today}

\author{Minjoon Park}
\address{
Department of Physics and MCTP, University of Michigan, \\
450 Church Street, Ann Arbor, MI 48109, USA
}
\ead{minjoonp@umich.edu}

\begin{abstract}
We consider the effect of quantum interactions on Pauli-Fierz massive gravity. With generic graviton cubic interactions, we observe that the 1-loop counterterms do not conform to the tree level structure of Pauli-Fierz action, resulting in the reappearance of the 6th mode ghost. Then to explore the quantum effects to the full extent, we calculate the resummed graviton propagator with an arbitrary interaction and analyze its complete structure, from which a minimal condition for the absence of the ghost is obtained.
\end{abstract}


\maketitle

\section{Introduction}
As a first step to reconcile gravity and quantum physics, or to understand the quantum nature of gravity, quantum corrections to general relativity(GR) have been extensively studied. The works on induced gravity\,\cite{Adler:1980bx} can be considered as an approach to matter loop corrections to Newton's constant, whereas 1-loop corrections to the massless graviton propagator were obtained for various types of fields: scalars\,\cite{Geist:1973my}, gauge bosons\,\cite{Capper:1974ed}, fermions\,\cite{Capper:1973mv}, and even gravitons themselves\,\cite{Capper:1973pv}. 
 
On a different front of the gravitational research, the theory of massive gravity has been an interesting topic for both theoretical and phenomenological reasons. On the theory side, it has been studied how to deal with the classical pathologies of massive gravity, the most famous one being the van Dam-Veltman-Zakharov(vDVZ) discontinuity\,\cite{vanDam:1970vg}:
Adding a mass term to the linearized Einstein-Hilbert action seems to be a natural way of giving mass to a graviton. 
But doing so breaks the general coordinate invariance(GCI) of GR, so that a massive theory ends up with
more degrees of freedom(DOFs) than the massless one. Requiring that none of these extra DOFs have any
pathology, one is forced to choose Pauli-Fierz(PF) theory\,\cite{Fierz:1939ix}.
Then the coupling of the extra scalar DOF to the sources remains finite even in the limit of vanishing graviton mass. 
That is, no matter how small the graviton mass is, the massive theory is finitely different from
the massless one. 

To elucidate, let us look at the linearized massive gravity action with a generic mass term
in a flat 4d background:
\bea\label{eqn:genericmassiveaction}
\fl S_{m_g,a} = \int\d^4 x \Big\{&\del_\alpha h^{\alpha\mu} \del_\beta h^\beta_\mu 
- \frac{1}{2}\del_\alpha h_{\mu\nu} \del^\alpha h^{\mu\nu}
+ \frac{1}{2} \del_\alpha h \del^\alpha h - \del_\alpha h^{\mu\alpha} \del_\mu h \nn\\
&- \frac{m_g^2}{2} (h^{\mu\nu} h_{\mu\nu} - a h^2) \Big\} \,,
\eea
with $h = \eta^{\mu\nu}h_{\mu\nu}$ and $\eta_{\mu\nu} = {\rm diag}(-1,1,1,1)$. The corresponding tree level propagator is
\be\label{eqn:genericmassiveprop}
\fl {\bf P}_{m_g,a}^{(0)}=\underbrace{\frac{i}{k^2+m_g^2} \Big(-\frac{{\bf I}_1}{2}+\frac{{\bf I}_2}{2}\Big)}_{H2} 
+ \underbrace{\frac{i}{k^2+m_g^2} \frac{{\bf I}_1}{6}}_{H0_1} 
+ \underbrace{\frac{-i}{k^2+\frac{4a-1}{2(1-a)}m_g^2} \frac{{\bf I}_1}{6}}_{H0_2} + \cdots\,,
\ee
where the ${\bf I}_i$'s are a complete set of tensor bases with 4 indicies, whose definitions will be given in \S \ref{sec-gloop}.
The propagator has a GR-like helicity-2 pole($H2$) and two helicity-0 ones($H0_1$ and $H0_2$), while $\cdots$ is terms that vanish upon contraction with conserved sources. In the $m_g\to0$ limit, (\ref{eqn:genericmassiveprop}) becomes
$
\frac{i}{k^2} \Big(-\frac{{\bf I}_1}{2}+\frac{{\bf I}_2}{2}\Big) + \cdots
$,
which is the same as the massless graviton propagator of GR, and hence we do not have any discontinuity problem.
But unfortunately (\ref{eqn:genericmassiveaction}) has a ghost DOF; (\ref{eqn:genericmassiveprop}) shows
that $H0_2$ has a negative coupling. In fact, it is this ghost that cancels the other scalar
in the massless limit, allowing a smooth transition to the massless theory.

By choosing $a=1$ the ghost mode decouples because its mass diverges, and we obtain PF theory: 
\be\label{eqn:pfmassiveprop}
{\bf P}_{m_g}^{(0)}=\frac{i}{k^2+m_g^2} \Big(-\frac{{\bf I}_1}{2}+\frac{{\bf I}_2}{2}\Big)
+ \frac{i}{k^2+m_g^2} \frac{{\bf I}_1}{6} + \cdots \,.
\ee
But then $H0_1$ survives the $m_g\to0$ limit, creating an untraversable gap between the massive and the massless theories. How or if we can remove this discontinuity has been an active subject of research\,\cite{Boulware:1973my}-\cite{vainshtein}.

The phenomenological reason to study massive gravity is a possibility of solving (a part of) the cosmological constant($\Lambda$) problem by modifying gravity at large distances. Among many proposed solutions to the $\Lambda$ problem, the infrared(IR) modification of gravity is the idea that gravity behaves differently at large distance scales compared to short distances. Let us assume that by some mechanism, \eg, \cite{Weinberg:1988cp}, we succeed in achieving vanishing $\Lambda$. The next step is to reconcile our zero $\Lambda$ with the small but non-zero $\Lambda$ calculated from the observational data.
The ``observed" $\Lambda$\,\cite{Riess:1998cb} is obtained with the assumption that GR holds at all scales. Then we can imagine that if the characteristic of the ``real" gravity is different from that of GR, interpreting the data with the right gravity may give an explanation to the accelerated expansion of the Universe without $\Lambda$.
But this idea gets severely constrained by experiments and observations which confirm the validity of GR from ${\rm mm}$ to the solar system scale. Therefore, the desired modification of gravity should be consistent with GR at short distances, while getting weaker than GR at large distances in order to mimic $\Lambda$. PF theory meets our demands in the IR, but fails to satisfy the short distance criterion because of the vDVZ discontinuity.

While interesting enough already at the classical level, not much attention has been paid to the quantum aspects of massive gravity theories. In this article, we will explore what the combination of ``quantum effects" and ``massive gravity" can offer, by investigating loop corrections to PF theory from the minimally coupled massive scalar(\S\ref{sec-sloop}), the graviton with a generic cubic interaction(\S\ref{sec-gloop}) and the graviton with an arbitrary interaction(\S\ref{sec-resum}).

\section{Effective action of a massive scalar}\label{sec-sloop}
As a warm-up exercise, let us consider the loop contribution from a real massive scalar field minimally coupled to gravity:
\be\label{eqn:gsa}
S_\phi = \frac{1}{2} \int \d^4x \,\sqrt{-g}\,(g^{\mu\nu}\partial_\mu\phi\partial_\nu\phi+m^2\phi^2)\,.
\ee
\begin{figure}
  \begin{center}
    \resizebox{10cm}{!}{\includegraphics{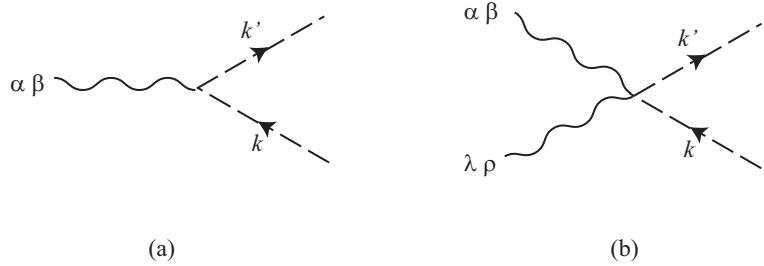}}
  \end{center}
  \caption{Graviton-scalar vertices.}
  \label{fig1}
\end{figure}
For $g_{\mu\nu} = \eta_{\mu\nu} + \frac{h_{\mu\nu}}{M_P}$, with $M_P$ the Planck mass,
we expand (\ref{eqn:gsa}) in $M_P^{-1}$ to get
\bea
\fl S_\phi &= \frac{1}{2} \int \d^4x &\Big( 1+ \frac{h}{2M_P} + \frac{h^2}{8M_P^2} 
- \frac{h^{\alpha\beta}h_{\alpha\beta}}{4M_P^2} \Big) \nn\\
\fl&& \Big\{ \Big( \eta^{\mu\nu} - \frac{h^{\mu\nu}}{M_P} + \frac{h^{\mu\rho} h^\nu_\rho}{M_P^2} \Big) 
\partial_\mu\phi\partial_\nu\phi+m^2\phi^2 \Big\} + {\cal O}(M_P^{-3}) \nn\\
\fl&= \frac{1}{2} \int \d^4x\, &\phi\Big[\; -\partial^2+m^2
+ \frac{h_{\alpha\beta}}{M_P} V^{(3)}_{\alpha\beta} 
+ \frac{h_{\alpha\beta}h_{\lambda\rho}}{M_P^2} V^{(4)}_{\alpha\beta;\lambda\rho}\Big]\phi
+ {\cal O}(M_P^{-3}) \,,
\eea
where $V^{(3)}$ and $V^{(4)}$ are the graviton-scalar-scalar vertex, figure \ref{fig1}(a),
and the 2-graviton-2-scalar vertex, figure \ref{fig1}(b), respectively, given in \ref{sloops}. 

Treating the graviton, $h_{\mu\nu}$, as an external field, we can calculate the scalar effective action, $W$;
\be
\fl e^{-i W[h]} = \int {\cal D}\phi\,e^{-i S_\phi[h]} 
= \Big[\det i \Big( -\partial^2+m^2 + \frac{h}{M_P} V^{(3)} + \frac{h h}{M_P^2} V^{(4)} + {\cal O}(M_P^{-3}) \Big) \Big]^{-1/2} ,\;
\ee
\ie,
\bea
\fl-i W[h] &=& -\frac{1}{2}\log\det i(-\partial^2+m^2) \nn\\
\fl&&-\frac{1}{2}\log\det\Big[1+\frac{1}{-\partial^2+m^2}
\Big(\frac{h}{M_P} V^{(3)} 
+ \frac{h h}{M_P^2} V^{(4)}\Big) + {\cal O}(M_P^{-3}) \Big] \nn\\
\fl&=& -\frac{1}{2}\tr\log i(-\partial^2+m^2) \nn\\
\fl&&+ \frac{1}{2}\sum_{n=1}^\infty \tr \frac{1}{n}\Big[{\rm P}_\phi \Big(\frac{\hat h}{M_P} i\hat V^{(3)}
+ \frac{\hat h \hat h}{M_P^2} i\hat V^{(4)}\Big) + {\cal O}(M_P^{-3}) \Big]^n ,
\eea 
where ${\rm P}_\phi = i/(p^2+m^2)$ is the scalar propagator in momentum space, $\hat A$ is a Fourier transform of $A$,
and we have suppressed indicies for simplicity. With
\be
{\rm P}_\phi \equiv \includegraphics[width=1cm]{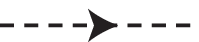}\,, \quad\hat h \equiv \includegraphics[width=1cm]{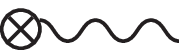}\,,
\ee
we write $W$ diagrammatically:
\bea\label{eqn:seffa}
\fl i W[h] &=& \frac{1}{2}\tr\log i(-\partial^2+m^2) \nn\\
\fl&&- \frac{1}{2}\Big[ \frac{1}{M_P}\includegraphics[bb=-5 15 80 50,keepaspectratio=true,width=1.6cm]{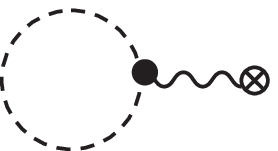}
+ \frac{1}{M_P^2}\includegraphics[bb=-5 15 70 50,keepaspectratio=true,width=1.4cm]{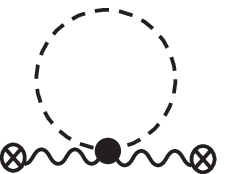} 
+ \frac{1}{2M_P^2}\includegraphics[bb=-5 15 110 50,keepaspectratio=true,width=2.2cm]{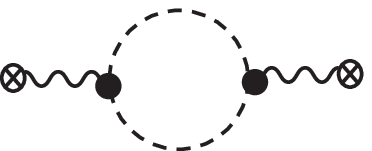} 
+ {\cal O}(M_P^{-3}) \Big] \nn\\
\fl&=& \frac{1}{2}\tr\log i(-\partial^2+m^2)
- \frac{1}{2} \Big[ \includegraphics[bb=-5 15 80 50,keepaspectratio=true,width=1.6cm]{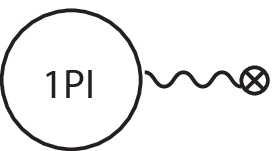} 
+ \includegraphics[bb=-5 15 110 50,keepaspectratio=true,width=2.2cm]{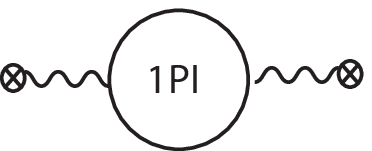} \nn\\
\fl&&\hspace{80pt} + {\rm 1PI's\; with\; three\; and\; more\; external\; graviton\; legs} \Big] \,,
\eea 
where
\bea\label{eqn:1gl1pi}
\fl\includegraphics[bb=0 15 80 50,keepaspectratio=true,width=1.6cm]{1ltp.eps}
&=& \frac{1}{M_P}\includegraphics[bb=-5 15 80 50,keepaspectratio=true,width=1.6cm]{13pv.eps} \nn\\
\fl&&+ {\rm higher\; loop\; contributions\; of\; {\cal O}({\it M_P^{-3}})\; with\; one\; external\; graviton\; leg} \nn\\
\fl&=& \int \d^4 x \frac{1}{\epsilon} \frac{im^4}{32\pi^2M_P} h + {\rm finite} + {\cal O}(M_P^{-3})\,, 
\eea
and
\bea\label{eqn:2gl1pi}
\fl\includegraphics[bb=0 15 110 50,keepaspectratio=true,width=2cm]{2l1pi.eps}
&=& \frac{1}{M_P^2}\includegraphics[bb=-5 15 70 50,keepaspectratio=true,width=1.4cm]{14pv.eps} 
+ \frac{1}{2M_P^2}\includegraphics[bb=-5 15 110 60,keepaspectratio=true,width=2.2cm]{23pv.eps} \nn\\
\fl&&+ {\rm higher\; loop\; contributions\; of\; {\cal O}({\it M_P^{-3}})\; with\; two\; external\; graviton\; legs} \nn\\
\fl&=& \int \d^4 x \frac{1}{\epsilon} \frac{i}{16\pi^2 M_P^2} \Big\{ \frac{1}{60}\sqrt{-g}(R^2+2R_{\mu\nu}R^{\mu\nu})\big|_{h^2} - \frac{m^2}{3} \sqrt{-g} R\big|_{h^2} \nn\\
\fl&&\hspace{80pt}+ \frac{m^4}{8}( h^2 - 2h_{\mu\nu}h^{\mu\nu}) \Big\} 
+ {\rm finite} + {\cal O}(M_P^{-3})\,.
\eea
Here we use dimensional regularization with $\epsilon = 4-d$, and $A\big|_{h^2}$ is the ${\cal O}(h^2)$ part of $A$. See Appendix A for details.

Combining (\ref{eqn:1gl1pi}) and (\ref{eqn:2gl1pi}) we get\,\cite{Geist:1973my}, 
\bea\label{eqn:sea}
\fl W &=& -\frac{1}{2\epsilon} \int \d^4x \sqrt{-g} \Big\{ \frac{m^4}{16\pi^2} - \frac{m^2}{48\pi^2 M_P^2}R + \frac{1}{960\pi^2M_P^2}(R^2+2R_{\mu\nu}R^{\mu\nu}) \Big\}\Big|_{h^2} \nn\\
\fl&&+ {\cal O}(M_P^{-3})\,,
\eea
where we used $\sqrt{-g} = 1+\frac{h}{2M_P}+\frac{h^2}{8M_P^2}-\frac{h_{\mu\nu}h^{\mu\nu}}{4M_P^2} + {\cal O}(M_P^{-3})$.
This only confirms the well known result\,\cite{'tHooft:1974bx}, \cite{Adler:1980bx} that the matter loop corrections take the form of
\be\label{eqn:mea}
\sim \int \d^4x \sqrt{-g} (\Lambda + a R + b_1 R^2 + b_2 R_{\mu\nu}R^{\mu\nu} + \cdots)\,.
\ee

Note that in obtaining (\ref{eqn:sea}) no reference to the specifics of the gravity sector was made, so that this result holds for massive gravity as well as for GR. That is, PF theory gets no more unusual or unexpected contributions from matter loops than GR does. Therefore, in order to see something interesting we need to consider graviton loops.

\section{Graviton loop corrections to PF}\label{sec-gloop}
The tree level quadratic action of PF massive gravity in a flat background is
\bea\label{eqn:pfa}
\fl S_{\rm PF} &=& \int\d^4 x \Big\{\del_\alpha h^{\alpha\mu} \del_\beta h^\beta_\mu 
- \frac{1}{2}\del_\alpha h_{\mu\nu} \del^\alpha h^{\mu\nu}
+ \frac{1}{2} \del_\alpha h \del^\alpha h - \del_\alpha h^{\mu\alpha} \del_\mu h \nn\\
\fl&&\hspace{35pt}- \frac{m_g^2}{2} (h^{\mu\nu} h_{\mu\nu} - h^2) \Big\} \nn\\
\fl&=& \frac{1}{2}\int \frac{\d^4 k}{(2\pi)^4} \hat h^{\mu\nu} 
\Big\{ (k^2+m_g^2)\Big({\bf I}_1 - \frac{{\bf I}_2}{2} \Big) 
+ \frac{{\bf I}_3}{2} - {\bf I}_4\Big\} \hat h^{\lambda\rho}\,,
\eea
from which we obtain the tree level massive graviton propagator,
\be\label{eqn:tlpfgp}
{\bf P}_{m_g}^{(0)} = \frac{i}{k^2+m_g^2}\Big( -\frac{{\bf I}_1}{3} + \frac{{\bf I}_2}{2} + \frac{{\bf I}_3}{2m_g^2}
- \frac{{\bf I}_4}{3m_g^2} + \frac{2{\bf I}_5}{3m_g^4} \Big) \,.
\ee
Here we introduce a complete set of tensor bases with 4 indicies:
\bea
\fl{\bf I}_1 = \eta_{\mu\nu} \eta_{\lambda\rho} \,,\quad
&{\bf I}_2 = \eta_{\mu\lambda}\eta_{\nu\rho} + \eta_{\mu\rho}\eta_{\nu\lambda} \,, \quad
{\bf I}_3 = \eta_{\mu\lambda} k_\nu k_\rho + \eta_{\mu\rho} k_\nu k_\lambda
+ (\mu \leftrightarrow \nu) \,, \nn\\
&{\bf I}_4 = \eta_{\mu\nu} k_\lambda k_\rho + k_\mu k_\nu \eta_{\lambda\rho} \,, \quad
{\bf I}_5 = k_\mu k_\nu k_\lambda k_\rho \,.
\eea
Note that we do not have to worry about fixing a gauge nor introducing Faddeev-Popov ghost, because the general covariance is explicitly broken by the graviton mass terms. To (\ref{eqn:pfa}), we add a generic cubic interaction:
\be\label{eqn:cubicint}
S_{\rm int} = \int \d^4 x \frac{\lambda}{(2!)^33!} \frac{m_g^2}{2M_P} ( \alpha h^{\mu_1}_{\nu_1} h^{\nu_1}_{\nu_2} h^{\nu_2}_{\mu_1} + \beta h^{\mu\nu} h_{\mu\nu} h + \gamma h^3)\,,
\ee
which gives a 3-graviton vertex:
\bea\label{eqn:3pv}
\fl V_g^{(3)} =& -\frac{i\lambda}{(2!)^33!} \frac{m_g^2}{2M_P} \big( \alpha \eta_{\mu_1\nu_3} \eta_{\nu_1\mu_2} \eta_{\nu_2\mu_3} + \beta \eta_{\mu_1\mu_2} \eta_{\nu_1\nu_2} \eta_{\mu_3\nu_3} + \gamma \eta_{\mu_1\nu_1} \eta_{\mu_2\nu_2} \eta_{\mu_3\nu_3} \nn\\
\fl &\hspace{70pt} + {\rm symmetrization\;in\;}\mu\nu + {\rm permutation\;in\;}123 \big) \,.
\eea

With the necessary building blocks ready, let us calculate loops. The details are given in \ref{gloops}. The first is the tadpole:
\be\label{eqn:tadpole}
\includegraphics[bb=0 15 80 50,keepaspectratio=true,width=1.6cm]{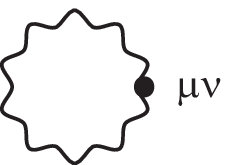}
=-\frac{1}{\epsilon}\frac{5i\lambda m_g^4}{192\pi^2M_P}(3\alpha+4\beta)\eta_{\mu\nu}+ {\rm finite}\,,
\ee
which is non-zero for a generic choice of the interaction, and therefore would imply that we have chosen the wrong background. But we can use the freedom of choosing $\alpha$, $\beta$ and $\gamma$, and 
\be
\beta = -\frac{3}{4}\alpha
\ee
renders (\ref{eqn:tadpole}) to vanish. With this choice, the vertex (\ref{eqn:3pv}) becomes
\be\label{eqn:fixedv3}
V_g^{(3)} = -i\lambda \frac{m_g^2}{16M_P}(\alpha {\bf J}_1 - \alpha {\bf J}_2 + 8\gamma {\bf J}_3)\,,
\ee
where ${\bf J}_{1(2,\,3)}$ is the symmetrization and permutation of $\eta_{\mu_1\nu_3} \eta_{\nu_1\mu_2} \eta_{\nu_2\mu_3}(\eta_{\mu_1\mu_2} \eta_{\nu_1\nu_2} \eta_{\mu_3\nu_3},$ $\eta_{\mu_1\nu_1} \eta_{\mu_2\nu_2} \eta_{\mu_3\nu_3})$.

Without the tadpole, the only 1-loop contribution to the 2-point function is
\bea\label{eqn:1l2pf}
\fl\includegraphics[bb=0 15 80 50,keepaspectratio=true,width=1.6cm]{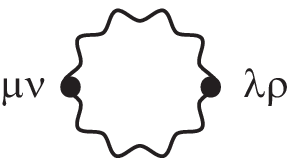}
\quad=&\frac{1}{\epsilon} \frac{5i \lambda^2m_g^2}{13824\pi^2M_P^2}\alpha^2\big\{(7k^2-m_g^2){\bf I}_1 - (5k^2-2m_g^2){\bf I}_2 + 18{\bf I}_3 - 18{\bf I}_4\big\} \nn\\
&+ {\cal O}(\frac{k^4}{M_P^2}) + {\rm finite} \,.
\eea
We can immediately see that it does not conform to (\ref{eqn:pfa}). Reconstructing the full counterterms out of (\ref{eqn:1l2pf}) reveals the trouble it causes:
\bea\label{eqn:2ptct}
\fl&&\hat h^{\mu\nu} \big\{ (7k^2 - m_g^2) {\bf I}_1 - (5k^2 - 2m_g^2) {\bf I}_2 + 18 {\bf I}_3 - 18 {\bf I}_4 \big\} \hat h^{\lambda\rho} \nn\\
\fl&& \Rightarrow\;\; 7\del_\alpha h \del^\alpha h
- 10\del_\alpha h_{\mu\nu} \del^\alpha h^{\mu\nu}
+ 72\del_\alpha h^{\alpha\mu} \del_\beta h^\beta_\mu 
- 36\del_\alpha h^{\mu\alpha} \del_\mu h
- m_g^2 (4h^{\mu\nu} h_{\mu\nu} - h^2) \nn\\
\fl&& \qquad= -33(\dot h_{00})^2 + \cdots \,.
\eea
As shown in \cite{Boulware:1973my} using the Arnowitt-Deser-Misner-decomposed version of (\ref{eqn:pfa}),
\bea\label{eqn:admpfa}
\fl S_{\rm PF} &=& \int\d^4x \;\Big\{ \pi^{ij} \dot{h_{ij}} - (\pi_{ij}^2 - \frac{1}{2}\pi_{ii}^2) + 2h_{0i} \del_j\pi^{ij}
+ \frac{1}{2}h_{00}(\del_i^2h_{ii}-\del_i\del_j h_{ij})+{}^3R|_{h^2} \nn\\
\fl&&\qquad\quad - \frac{m_g^2}{4}(h_{ij}^2- h_{ii}^2-2h_{0i}{}^2+2h_{00}h_{ii}) \Big\} \,,
\eea
with $\pi_{ij}$ the conjugate momentum to $h_{ij}$ and ${}^3R$ the curvature scalar constructed with $h_{ij}$, 
PF theory comes to have 5 healthy DOFs because among the 10 DOFs of $h_{\mu\nu}$, $h_{00}$ and $h_{0i}$ are non-dynamical, and furthermore the action is linear in $h_{00}$. That is, being a Lagrange multiplier, $h_{00}$ provides a constraint to eliminate another DOF, the 6th ghost mode. What we see from (\ref{eqn:2ptct}) is that quantum effects remove such a feature, incurring the reappearance of the ghost DOF. 

Similar outcome is obtained with the 3-point function:
\bea\label{eqn:1l3pf}
\fl\includegraphics[bb=0 30 80 65,keepaspectratio=true,width=1.6cm]{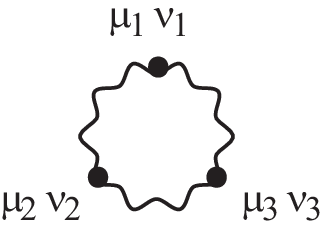}
\quad=&\frac{1}{\epsilon}\frac{5i\lambda^3m_g^4}{442368\pi^2M_P^3}\alpha^2\Big\{ 13\alpha{\bf J}_1 - \frac{1}{3}(35\alpha+32\gamma){\bf J}_2 + (11\alpha+16\gamma){\bf J}_3 \Big\} \nn\\
&+ {\rm terms\; with\; external\; momenta} + {\rm finite} \,.
\eea
No nontrivial choice of $\alpha$ and $\gamma$ can make (\ref{eqn:1l3pf}) conform to the tree level vertex (\ref{eqn:fixedv3}).

If we interpret terms with derivatives in (\ref{eqn:pfa}) as $2M_P^2\sqrt{-g}R\big|_{h^2}$ with $g_{\mu\nu}=\eta_{\mu\nu}+\frac{h_{\mu\nu}}{M_P}$, we might well have to consider derivative interactions such as $2M_P^2\sqrt{-g}R\big|_{h^3}$ as well as (\ref{eqn:cubicint}). The case of a more general interaction including the cubic and quartic expansion of $\sqrt{-g}R$ is analyzed in \ref{ggloops} with the same conclusion.

\section{Resummed graviton propagator}\label{sec-resum}
We just saw the breakdown of PF theory at the quantum level. But since we tried only one type of interaction, it is still possible that the situation gets better for different types of well designed interactions. To explore the full capability of quantum effect, let us perform a general analysis independent of the details of the interactions.

No matter what the interaction is, the one particle irreducible diagram(1PI), $\bf\Pi$, for the 2-point function may be written as
\be\label{eqn:1pi}
{\bf\Pi}_{\mu\nu;\lambda\rho} = \sum_{j=1}^5 b_j(k) {\bf I}_j \,,
\ee
because ${\bf I}_j$ is a complete basis. In the previous section, we concentrated only on ${\mathcal O}(k^2)$ part of $\bf\Pi$, but of course there are pieces of ${\mathcal O}(k^4)$ and higher. Being generically nonrenormalizable, the tree level PF theory cannot handle the divergences of higher powers of $k$. Here we take the idea of \cite{Donoghue:1994dn} and treat the massive gravity as an effective field theory(EFT). That is, we assume the tree level PF action gets complemented by EFT terms, which can absorb divergences from the loops. But regardless of the EFT treatment, ${\mathcal O}(k^0)$ and ${\mathcal O}(k^2)$ divergencies must be absorbed by the bare theory. Then, the very first requirement for our interaction is
\bea\label{eqn:condiv}
&&{\mathcal O}(k^0)\; {\rm and}\; {\mathcal O}(k^2)\; {\rm divergencies\; of\; 1PI\; for\; the\; 2{\scriptstyle -}point\; function} \nn\\
&&{\rm conform\; to\; the\; tree\; level\; PF\; action.}
\eea
From here on, it is understood that $b_i$'s are the finite parts of the quantum corrections with the divergences taken care of.

To obtain the resummed graviton propagator 
\be\label{eqn:pfrsprops}
{\bf P}_{m_g} = {\bf P}_{m_g}^{(0)} + {\bf P}_{m_g}^{(0)} \cdot {\bf\Pi} \cdot {\bf P}_{m_g}^{(0)} 
+ {\bf P}_{m_g}^{(0)} \cdot ({\bf\Pi} \cdot {\bf P}_{m_g}^{(0)})^2 + \cdots \,,
\ee
we need to find $({\bf\Pi} \cdot {\bf P}_{m_g}^{(0)})^n$ as a function of $n$. By writing
\be
{\bf\Pi} \cdot {\bf P}_{m_g}^{(0)} = \sum_i r_i {\bf I}_i\,,
\ee 
and then
\be
({\bf\Pi} \cdot {\bf P}_{m_g}^{(0)})^{n+1} = \sum_i a_i^{(n)} {\bf I}_i
= \sum_i r_i {\bf I}_i \cdot \sum_j a_j^{(n-1)} {\bf I}_j \,,
\ee
we can obtain the recurrence relations between $a_i^{(n)}$ and $a_j^{(n+1)}$. Explicit solutions for $a_i^{(n)}$ are given in \ref{resum}, and the resummed propagator is
\bea\label{eqn:pfloopsum}
\fl {\bf P}_{m_g} &=& {\bf P}_{m_g}^{(0)} 
+ \sum_{n=0}^\infty{\bf P}_{m_g}^{(0)} \cdot ({\bf\Pi} \cdot {\bf P}_{m_g}^{(0)})^{n+1} \nn\\
\fl &=& - \frac{i}{k^2+m_g^2+2b_2}\frac{{\bf I}_1}{3} + \frac{i}{k^2+m_g^2+2b_2} \frac{{\bf I}_2}{2} - i \frac{b_5k^4+2(2b_3+b_4)k^2+b_1+2b_2}{d_{m_g}\hspace{-2pt}(k)} \frac{{\bf I}_1}{3} \nn\\
\fl&&+ ({\bf I}_3 + {\bf I}_4 + {\bf I}_5) \,,
\eea
where
\bea
\fl d_{m_g}\hspace{-2pt}(k) &=& 2b_5k^6 -(4b_1+2b_2-3m_g^2)(2b_2+m_g^2) \nn\\
\fl&&+\big\{8b_3+b_4(4+3b_4)-b_5(3b_1+2b_2)+2b_5 m_g^2\big\}k^4 \nn\\
\fl&&+\big\{2b_1(1-6b_3)+4b_2(1-2b_3-b_4)+2(4b_3-b_4)m_g^2\big\}k^2 \,.
\eea
Comparing the first two terms of (\ref{eqn:pfloopsum}) with the tree level propagator, (\ref{eqn:tlpfgp}), we can immediately see that the tree level theory gets renormalized by $b_2$, which is a progress over the 1-loop analysis where it was not clear which counterterm renormalized what. Next, by rewriting (\ref{eqn:pfloopsum}) as
\bea\label{eqn:pfloopsumrw}
\fl {\bf P}_{m_g}&=& \frac{i}{k^2+m_g^2+2b_2} \Big(- \frac{{\bf I}_1}{2} + \frac{{\bf I}_2}{2}\Big) + \frac{i}{k^2+m_g^2+2b_2} \frac{{\bf I}_1}{6} + \frac{-i}{k^2 + {\mathcal M}^2} \frac{{\bf I}_1}{6} \nn\\
\fl&&+ ({\bf I}_3 + {\bf I}_4 + {\bf I}_5) \,,
\eea
with
\bea
\fl{\mathcal M}^2 &=& \frac{1}{2\{b_5k^4+2(2b_3+b_4)k^2+b_1+2b_2\}}\big[\{3b_4^2-b_5(3b_1+2b_2-2m_g^2)\}k^4 \nn\\
\fl&&\qquad\qquad - 2\{6b_1b_3+2b_2(2b_3+b_4)-m_g^2(4b_3-b_4)\}k^2 \nn\\
\fl&&\qquad\qquad - (4b_1+2b_2-3m_g^2)(2b_2+m_g^2)\big] \,,
\eea
we notice that (\ref{eqn:pfloopsumrw}) takes the same form as the tree level \emph{non}-PF propagator, (\ref{eqn:genericmassiveprop}), implying that a generic interaction revives the ghost 6th mode. But (\ref{eqn:pfloopsum}) shows how to avoid this pathology: The ghost pole can be removed if
\be\label{eqn:ngcon}
b_5k^4+2(2b_3+b_4)k^2+b_1+2b_2 = 0\,.
\ee
Therefore, in order to be quantum-safe, 1PI from a desirable interaction should satisfy at least (\ref{eqn:condiv}) and (\ref{eqn:ngcon}).

\section{Discussion}
Through straightforward loop calculations and the analysis on the propagator, we showed that although loop corrections from a quantum interaction may spoil PF massive gravity by reviving the 6th mode ghost, we may still be able to have a healthy theory by requiring the allowed interactions to satisfy appropriate conditions.

Then the next task would be to find the right interactions. For this purpose, understanding why the cubic interaction of \S\ref{sec-gloop} failed would be useful. In fact, the results of \S\ref{sec-gloop} should not be surprising, once we realize that in PF the elimination of the 6th mode is achieved by an onshell symmetry. To identify this symmetry, we start with the action, (\ref{eqn:genericmassiveaction}), with a generic mass term. The equation of motion for the graviton, $h$, with a source, $T$, is
\bea
T_{\mu\nu} &=& \partial^2 h_{\mu\nu} -\eta_{\mu\nu}\partial^2 h - \partial_\mu\partial_\alpha h^\alpha_\nu
- \partial_\nu\partial_\alpha h^\alpha_\mu + \partial_\mu\partial_\nu h + \eta_{\mu\nu}\partial_\alpha\partial_\beta h^{\alpha\beta} \nn\\
&&- m_g^2(h_{\mu\nu} - a \eta_{\mu\nu} h) \,. 
\eea
On the RHS, terms with derivatives come from the Einstein tensor, and the Bianchi identity guarantees
that their contraction with $\partial^\mu$ vanishes. Then for a conserved source, \ie, $\partial^\mu T_{\mu\nu} = 0$,
we get the following onshell constraint:
\be\label{eqn:osc}
0 = \partial^\mu h_{\mu\nu} - a \partial_\nu h \,.
\ee
Next we vary (\ref{eqn:genericmassiveaction}) under the infinitesimal coordinate transformation, $x \to x+\xi$ or $h_{\mu\nu} \to h'_{\mu\nu} = h_{\mu\nu} + \partial_\mu\xi_\nu + \partial_\nu\xi_\mu$. Again, terms with derivatives are $\sqrt{-g} R\big|_{h^2}$ and therefore invariant. A remainder of the variation of the mass term is
\be
\delta S_{m_g} = m_g^2 \int\d^4x \,\xi^\nu (\partial^\mu h_{\mu\nu} - a \partial_\nu h) \,.
\ee
Then a generic massive gravity seems to have GCI if the constraint (\ref{eqn:osc}) is imposed.
Of course this is not true, because we have yet to take into account that the transformed field should also satisfy (\ref{eqn:osc}), \ie,
\be\label{eqn:cstrntonxi}
0 = \partial^\mu h'_{\mu\nu} - a \partial_\nu h'
= (1-2a)\partial_\nu \partial_\mu\xi^\mu + \partial^2\xi_\nu \,.
\ee
Decomposing $\xi$ into a transverse vector $\xi^{\rm T}$ and a longitudinal scalar $\sigma$ such that
$\xi_\mu = \xi^{\rm T}_\mu + \partial_\mu\sigma$, (\ref{eqn:cstrntonxi}) becomes
\be\label{eqn:cstrntoxi1}
\partial^2\xi^{\rm T}_\nu + 2(1-a)\partial_\nu\partial^2\sigma = 0\,.
\ee
Thus we end up with $\xi^{\rm T} = 0$ and for $a \neq 1$ $\sigma$ should also vanish: Non-PF action has no symmetry. 
But when $a=1$, which is the case of PF theory, (\ref{eqn:cstrntoxi1}) can be satisfied with
a nontrivial $\sigma$, and PF theory has a residual symmetry parametrized by $\xi_\mu = \partial_\mu\sigma$.
Since this symmetry works under the \emph{onshell} constraint (\ref{eqn:osc}), it may not be preserved when we go offshell in the loop calculations. 

Therefore further efforts to find a quantum-safe theory of massive gravity can be directed in two different ways:
\begin{enumerate}
\item We may try to construct a nonlinear completion of PF where the 6th mode is removed by a \emph{full} symmetry. 
\item We can attempt to directly find a quantum interaction whose 1PI satisfies \eg, (\ref{eqn:condiv}) and (\ref{eqn:ngcon}).
\end{enumerate}
With various versions\,\cite{Damour:2002ws} of the completion of PF already at hand, it would be straightforward to pursue (i), which in the end might lead us to the right interaction sought after in (ii).

\ack
We thank Ratindranath Akhoury, Henriette Elvang, Alberto Iglesias, Nemanja Kaloper, Choonkyu Lee, James Liu, Inyong Park and Scott Watson for enlightening discussions.

\appendix
\section{Massive scalar loops}\label{sloops}
From
\bea
\fl S_\phi &=& \int \d^4x \Big( 1+ \frac{h}{2M_P} + \frac{h^2}{8M_P^2} 
- \frac{1}{4M_P^2}h^{\alpha\beta}h_{\alpha\beta} \Big) \nn\\
\fl&&\qquad \Big\{ \big( \eta^{\mu\nu} - \frac{h^{\mu\nu}}{M_P} + \frac{h^{\mu\rho} h^\nu_\rho}{M_P^2} \big) 
\partial_\mu\phi\partial_\nu\phi+m^2\phi^2 \Big\} 
+ {\cal O}(M_P^{-3}) \nn\\
\fl&=& \int \d^4x \Big[\; \eta^{\mu\nu}\partial_\mu\phi\partial_\nu\phi+m^2\phi^2 \nn\\
\fl&&+ \frac{h_{\alpha\beta}}{M_P} \Big\{ 
- \frac{1}{2}(\eta^{\alpha\mu}\eta^{\beta\nu} + \eta^{\alpha\nu}\eta^{\beta\mu}) \partial_\mu\phi^*\partial_\nu\phi 
+ \frac{1}{2}\eta^{\alpha\beta}(\partial^\mu\phi\partial_\mu\phi + m^2 \phi^2) \Big\} \nn\\
\fl&&+ \frac{h_{\alpha\beta}h_{\lambda\rho}}{8M_P^2} 
\Big\{ \Big(\,\eta^{\alpha\mu}(\eta^{\lambda\nu}\eta^{\beta\rho} + \eta^{\rho\nu}\eta^{\beta\lambda})
+ \eta^{\alpha\nu}(\eta^{\lambda\mu}\eta^{\beta\rho} + \eta^{\rho\mu}\eta^{\beta\lambda}) 
+ (\alpha \leftrightarrow \beta) \,\Big) \,\partial_\mu\phi\partial_\nu\phi \nn\\
\fl&&\quad\qquad - \Big(\,\eta^{\alpha\beta}(\eta^{\lambda\mu}\eta^{\rho\nu}+\eta^{\lambda\nu}\eta^{\rho\mu})
+ \eta^{\lambda\rho}(\eta^{\alpha\mu}\eta^{\beta\nu}+\eta^{\alpha\nu}\eta^{\beta\mu}) \,\Big) \,
\partial_\mu\phi\partial_\nu\phi \nn\\
\fl&&\quad\qquad+ (\eta^{\alpha\beta}\eta^{\lambda\rho} - \eta^{\alpha\lambda}\eta^{\beta\rho} 
- \eta^{\alpha\rho}\eta^{\beta\lambda}) (\partial^\mu\phi\partial_\mu\phi+m^2\phi^2) \Big\} \Big] 
+ {\cal O}(M_P^{-3}) \,,
\eea
we can read off a graviton-scalar-scalar vertex, $V^{(3)}$,
\be\label{eqn:gssv}
i \hat V^{(3)}_{\alpha\beta} = \frac{i}{2} \Big\{ -(k'_\alpha k_\beta + k'_\beta k_\alpha) 
+ \eta_{\alpha\beta} (k'\cdot k+m^2) \Big\}\,,
\ee
and a 2-graviton-2-scalar vertex, $V^{(4)}$,
\bea\label{eqn:ggssv}
i\hat V^{(4)}_{\alpha\beta;\lambda\rho} &=& \frac{i}{8}
\Big[\, \{ \eta_{\alpha\lambda}(k'_\beta k_\rho + k'_\rho k_\beta)
+ \eta_{\alpha\rho}(k'_\beta k_\lambda + k'_\lambda k_\beta) + (\alpha \leftrightarrow \beta) \} \nn\\
&&\quad - \{ \eta_{\alpha\beta}(k'_\lambda k_\rho + k'_\rho k_\lambda)
+ \eta_{\lambda\rho}(k'_\alpha k_\beta + k'_\beta k_\alpha) \} \nn\\
&&\quad + (\eta_{\alpha\beta}\eta_{\lambda\rho} 
- \eta_{\alpha\lambda}\eta_{\beta\rho} - \eta_{\alpha\rho}\eta_{\beta\lambda})(k'\cdot k + m^2) \Big] \,.
\eea
Then the one loop linear tadpole diagram with the dimensional regularization is
\bea
\fl\frac{1}{M_P}\includegraphics[bb=-5 15 80 50,keepaspectratio=true,width=1.6cm]{13pv.eps}
= \frac{\hat h(0)}{M_P}\int \frac{\d^4 p}{(2\pi)^4}\frac{i}{p^2+m^2}\,i\hat V^{(3)}\nn\\
\fl\qquad= -\eta_{\alpha\beta}\frac{\hat h_{\alpha\beta}(0)}{2M_P}\int \frac{\d^d p}{(2\pi)^d}
\frac{(1-\frac{2}{d})p^2+m^2}{p^2+m^2}
= \int \d^4 x \frac{1}{\epsilon} \frac{im^4}{32\pi^2M_P} h + {\rm finite}\,,
\eea
where we use $d=4-\epsilon$,
\be
\fl\int \frac{\d^d l}{(2\pi)^d}\frac{l^{2u}}{(l^2+\Delta)^n} = \frac{i}{(4\pi)^{d/2}\Delta^{n-u-d/2}}\frac{\Gamma(u+d/2)\Gamma(n-u-d/2)}{\Gamma(d/2)\Gamma(n)}\,,
\ee
and 
\be
\fl\qquad\qquad\hat h(0) = \int \d^4 k \,\delta^4(k)\, \hat h(k) = \int \d^4 k \int \frac{\d^4 x}{(2\pi)^4} \,e^{ik\cdot x} \hat h(k)
= \int \d^4 x \,h(x)\,.
\ee 
Similarly, between the two 1-loop diagrams with two external graviton legs
the simpler one is 
\bea
\fl\frac{1}{M_P^2}\includegraphics[bb=0 15 70 50,keepaspectratio=true,width=1.4cm]{14pv.eps} 
&=& \int \frac{\d^4 k}{(2\pi)^4}\frac{\hat h(k) \hat h(-k)}{M_P^2}
\int \frac{\d^4 p}{(2\pi)^4}\frac{i}{p^2+m^2}\,i\hat V^{(4)} \nn\\
\fl&=& -\int \frac{\d^4 k}{(2\pi)^4} \frac{\hat h_{\alpha\beta}(k) \hat h_{\lambda\rho}(-k)}{8M_P^2}
\int \frac{\d^d p}{(2\pi)^d} \frac{(1-\frac{4}{d})p^2+m^2}{p^2+m^2}({\bf I}_1-{\bf I}_2) \nn \\
\fl&=& \int \d^4 x \frac{1}{\epsilon} \frac{im^4}{64\pi^2M_P^2}
(h^2-2h_{\mu\nu}h^{\mu\nu}) + {\rm finite} \,.
\eea
The more complicated diagram is
\bea\label{eqn:23pvint}
\fl&&\frac{1}{M_P^2}\includegraphics[bb=-5 15 110 60,keepaspectratio=true,width=2.2cm]{23pv.eps} \nn\\
\fl&&= \int \frac{\d^4 k}{(2\pi)^4} \frac{\hat h(k) \hat h(-k)}{M_P^2}
\int \frac{\d^4 p}{(2\pi)^4}\frac{i}{p^2+m^2}i\hat V^{(3)} \frac{i}{(p+k)^2+m^2}i\hat V^{(3)} \nn\\
\fl&&= \int \frac{\d^4 k}{(2\pi)^4} \frac{\hat h(k) \hat h(-k)}{4M_P^2}
\int_0^1\d x \int \frac{\d^d l}{(2\pi)^d} \frac{1}{(l^2+m^2+x(1-x)k^2)^2} \nn\\
\fl&&\quad\Big[ \Big\{ \frac{4}{d(d+2)}({\bf I}_1+{\bf I}_2) + \Big( 1-\frac{4}{d} \Big){\bf I}_1 \Big\}l^4 \\
\fl&&\quad+ \Big\{ \frac{(1-2x)^2}{d}(k^2{\bf I}_1+{\bf I}_3-2{\bf I}_4) 
+ \Big( 1-\frac{2}{d} \Big)\Big( 2\big(m^2-x(1-x)k^2\big){\bf I}_1+2x(1-x){\bf I}_4 \Big) \Big\}l^2 \nn\\
\fl&&\quad+\big( m^2-x(1-x) k^2 \big)^2{\bf I}_1
+2x(1-x)\Big( \big(m^2-x(1-x)k^2 \big){\bf I}_4+2x(1-x){\bf I}_5 \Big) \Big] \,,\nn
\eea
where $l = p+x k$ and $x$ is a Feynman parameter. Its $\epsilon^{-1}$ part is
\bea
\fl&&\int \frac{\d^4 k}{(2\pi)^4} \,\hat h(k) \hat h(-k) 
\frac{1}{\epsilon}\frac{i}{16\pi^2 M_P^2}\Big\{ \frac{1}{20}\Big( k^4 {\bf I}_1 + \frac{1}{6}k^4 {\bf I}_2
- \frac{1}{6}k^2 {\bf I}_3 - k^2 {\bf I}_4 + \frac{4}{3}{\bf I}_5 \Big) \nn\\
\fl&&\hspace{120pt}
- \frac{m^2}{6}\Big( k^2 {\bf I}_1 - \frac{1}{2}k^2{\bf I}_2 + \frac{1}{2}{\bf I}_3 - {\bf I}_4 \Big)
- \frac{m^4}{4}( {\bf I}_1 - {\bf I}_2) \Big\} \nn\\
\fl&&= \int \d^4 x \frac{1}{\epsilon} \frac{i}{16\pi^2 M_P^2} \Big\{ - \frac{2m^2}{3} \sqrt{-g}R\big|_{h^2} + \frac{1}{30}\sqrt{-g}(R^2+2R_{\mu\nu}R^{\mu\nu})\big|_{h^2} \nn\\
\fl&&\hspace{95pt} - \frac{m^4}{4}( h^\mu_\mu{}^2 - 2h_{\mu\nu}h^{\mu\nu}) \Big\}\,,
\eea
with $A\big|_{h^2}$ the ${\cal O}(h^2)$ part of $A$.

\section{Graviton loops}\label{gloops}
\begin{figure}
  \begin{center}
    \resizebox{11cm}{!}{\includegraphics{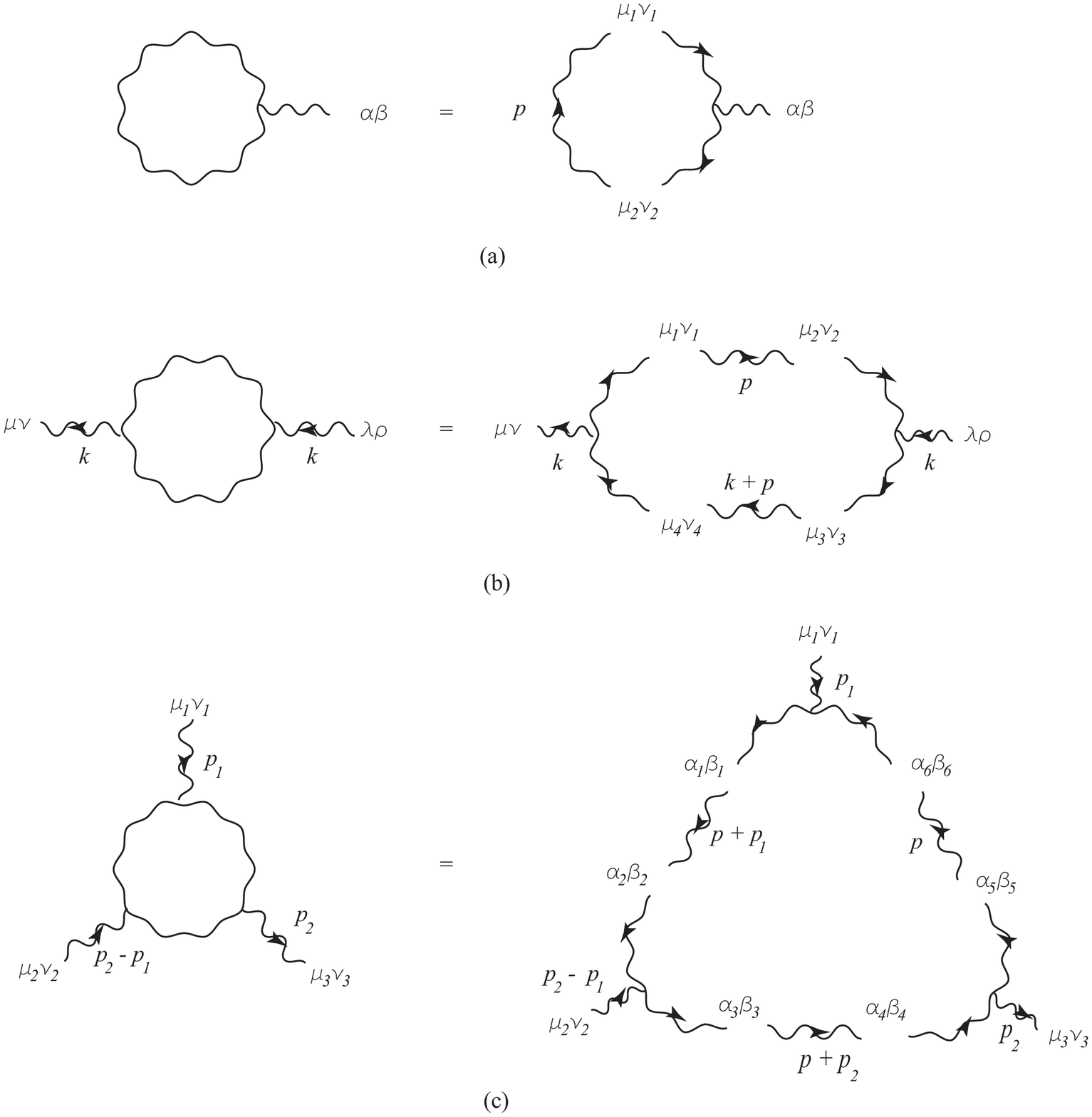}}
  \end{center}
  \caption{Graviton 1-loops.}
  \label{g1loops}
\end{figure}
All the 1-loop diagrams constructed with the 3-graviton vertex $V_g^{(3)}$, (\ref{eqn:3pv}),
are drawn in figure \ref{g1loops}. Evaluating them is straightforward but laborious and tedious. The tadpole diagram,
\bea
\fl{\rm (a)} &=& \int \frac{\d^4p}{(2\pi)^4} V_g^{(3)}(\alpha\beta;\mu_1\nu_1;\mu_2\nu_2) 
{\bf P}_{m_g}^{(0)}{}_{\mu_1\nu_1;\mu_2\nu_2}(p) \,, \nn\\
\fl&=& \frac{\lambda m_g^2}{M_p}\eta_{\mu\nu}\int\frac{\d^dp}{(2\pi)^d} \frac{1}{p^2+m_g^2}\Big\{ \frac{3\alpha+(d+2)\beta+3d\gamma}{9m_g^4d}p^4 \nn\\
\fl&&\qquad\qquad+ \frac{(3d+1)\alpha+(d^2-d+4)\beta-2(d-3)d\gamma}{6} \Big(  \frac{p^2}{m_g^2d} + \frac{1}{2} \Big) \Big\} \nn\\
\fl&=& -\frac{1}{\epsilon}\frac{5i\lambda m_g^4}{192\pi^2M_P}(3\alpha+4\beta)\eta_{\mu\nu}+ {\rm finite}\,,
\eea
fixes $\beta=-\frac{3}{4}\alpha$. Then
\bea\label{eqn:2ptftnint}
\fl{\rm (b)} &=& \int \frac{\d^4p}{(2\pi)^4} V_g^{(3)}(\mu\nu;\mu_1\nu_1;\mu_4\nu_4) {\bf P}_{m_g}^{(0)}{}_{\mu_1\nu_1;\mu_2\nu_2}(p)
V_g^{(3)}(\lambda\rho;\mu_2\nu_2;\mu_3\nu_3) {\bf P}_{m_g}^{(0)}{}_{\mu_3\nu_3;\mu_4\nu_4}(p+k)  \nn\\
\fl&=& -\frac{\lambda^2m_g^4}{M_P^2}\int\frac{\d^dl}{(2\pi)^d}\int_0^1\d x\frac{1}{\{l^2+m_g^2+x(1-x)k^2\}^2} \nn\\
\fl&& \Big[ \Big( \frac{7\alpha^2+32\alpha\gamma-128\gamma^2}{576m_g^2}\big(1-2x(1-x)\big)k^2 + \frac{(\alpha-2\gamma)(\alpha+4\gamma)}{18} \Big) {\bf I}_1 \nn\\
\fl&& - \frac{\alpha^2}{8}\Big( \frac{3\big( 1-2x(1-x) \big)}{8m_g^2}k^2 + 1 \Big) {\bf I}_2
- \frac{\alpha^2}{32m_g^2}\big( 1-2x(1-x) \big)({\bf I}_3-{\bf I}_4) \nn\\
\fl&& + \Big\{ \Big( \frac{\big( 25-58x(1-x)\big)\alpha^2-32\big(5-7x(1-x) \big)\alpha\gamma + 128\big(5-9x(1-x) \big)\gamma^3}{1152m_g^2}k^2 \nn\\
\fl&&\qquad\quad + \frac{(\alpha-2\gamma)(\alpha+4\gamma)}{18} \Big) {\bf I}_1 \nn\\
\fl&&\qquad - \frac{\alpha^2}{8}\Big( \frac{5-16x(1-x)}{16m_g^2}k^2 + 1 \Big) {\bf I}_2
- \frac{\alpha^2}{192m_g^2}\big( 5-16x(1-x) \big){\bf I}_3 \nn\\
\fl&&\qquad + \frac{\big(15-56x(1-x)\big)\alpha^2+64x(1-x)\alpha\gamma}{576m_g^2}{\bf I}_4) \Big\} \frac{l^2}{m_g^2} \nn\\
\fl&& + \Big\{ \Big( - \frac{\big( 17-50x(1-x)\big)\alpha^2+64\big(5-8x(1-x) \big)\alpha\gamma - 1536\big(1-3x(1-x) \big)\gamma^3}{3456m_g^2}k^2 \nn\\
\fl&&\qquad\quad + \frac{19\alpha^2-48\alpha\gamma+192\gamma^2}{576} \Big) {\bf I}_1 \nn\\
\fl&&\qquad + \frac{\alpha^2}{18}\Big( \frac{1+22x(1-x)}{64m_g^2}k^2 - 1 \Big) {\bf I}_2
- \frac{\alpha^2}{576m_g^2}\big( 21-50x(1-x) \big){\bf I}_3 \nn\\
\fl&&\qquad + \frac{\big(71-230x(1-x)\big)\alpha^2-64\big(1-10x(1-x)\big)\alpha\gamma}{1728m_g^2}{\bf I}_4) \Big\} \frac{l^4}{m_g^4} \nn\\
\fl&& + \Big\{ \Big( - \frac{65(1-2x)^2\alpha^2-64\big(11-45x(1-x) \big)\alpha\gamma + 768\big(3-10x(1-x) \big)\gamma^3}{6912m_g^2}k^2 \nn\\
\fl&&\qquad\quad + \frac{5\alpha^2-32\alpha\gamma+128\gamma^2}{576} \Big) {\bf I}_1 \nn\\
\fl&&\qquad + \frac{\alpha^2}{96}\Big( \frac{7+20x(1-x)}{72m_g^2}k^2 - 1 \Big) {\bf I}_2
- \frac{\alpha^2}{3456m_g^2}\big( 19-60x(1-x) \big){\bf I}_3 \nn\\
\fl&&\qquad + \frac{5\big(7-36x(1-x)\big)\alpha^2-64\big(2-15x(1-x)\big)\alpha\gamma}{3456m_g^2}{\bf I}_4) \Big\} \frac{l^6}{m_g^6} \nn\\
\fl&&- \frac{(\alpha^2-24\alpha\gamma+96\gamma^2){\bf I}_1 + \alpha^2{\bf I}_2}{864}\frac{l^8}{m_g^8} + {\mathcal O}(k^4) \Big] \nn\\
\fl&=& \frac{\alpha^2}{\epsilon} \frac{5i \lambda^2m_g^2}{13824\pi^2M_P^2}\big\{(7k^2-m_g^2){\bf I}_1 - (5k^2-2m_g^2){\bf I}_2 + 18{\bf I}_3 - 18{\bf I}_4\big\} 
+ {\cal O}\Big(\frac{k^4}{M_P^2}\Big) + {\rm finite} \,,\nn\\
\fl&&\\
\label{eqn:3ptftnint}
\fl{\rm (c)} &=& \int \frac{\d^4p}{(2\pi)^4} V_g^{(3)}(\mu_1\nu_1;\alpha_1\beta_1;\alpha_6\beta_6) {\bf P}_{m_g}^{(0)}{}_{\alpha_1\beta_1;\alpha_2\beta_2}(p+p_1) \nn\\
\fl&&\quad V_g^{(3)}(\mu_2\nu_2;\alpha_2\beta_2;\alpha_3\beta_3) {\bf P}_{m_g}^{(0)}{}_{\alpha_3\beta_3;\alpha_4\beta_4}(p+p_2) 
V_g^{(3)}(\mu_3\nu_3;\alpha_4\beta_4;\alpha_5\beta_5) {\bf P}_{m_g}^{(0)}{}_{\alpha_5\beta_5;\alpha_6\beta_6}(p) \nn\\
\fl&=& \frac{\lambda^3m_g^6}{256M_P^3} \int \frac{\d^dl}{(2\pi)^d} \frac{1}{(l^2+m_g^2)^3} 
\Big\{ \alpha^3 {\bf J}_1 - \alpha^3 {\bf J}_2 + \frac{31\alpha^3-96\alpha^2\gamma+768\alpha\gamma^2-2048\gamma^3}{27}{\bf J}_3 \nn\\
\fl&& + \Big( \frac{3\alpha^3}{2m_g^2} {\bf J}_1 - \frac{3\alpha^3}{2m_g^2} {\bf J}_2 + \frac{31\alpha^3-96\alpha^2\gamma+768\alpha\gamma^2-2048\gamma^3}{18m_g^2}{\bf J}_3 \Big) l^2 \nn\\
\fl&& + \Big( \frac{7\alpha^3}{6m_g^4} {\bf J}_1 - \frac{67\alpha^3-32\alpha^2\gamma}{54m_g^4} {\bf J}_2 + \frac{21\alpha^3+32\alpha^2\gamma-384\alpha\gamma^2+1024\gamma^3}{18m_g^4}{\bf J}_3 \Big) l^4 \nn\\ 
\fl&& + \Big( \frac{5\alpha^3}{6m_g^6} {\bf J}_1 - \frac{28\alpha^3-44\alpha^2\gamma}{27m_g^6} {\bf J}_2 + \frac{101\alpha^3+264\alpha^2\gamma-4224\alpha\gamma^2+11264\gamma^3}{108m_g^6}{\bf J}_3 \Big) l^6 \nn\\ 
\fl&& + \Big( \frac{13\alpha^3}{36m_g^8} {\bf J}_1 - \frac{181\alpha^3-512\alpha^2\gamma}{324m_g^8} {\bf J}_2 + \frac{77\alpha^3-400\alpha^2\gamma+1152\alpha\gamma^2-3072\gamma^3}{108m_g^8}{\bf J}_3 \Big) l^8 \nn\\ 
\fl&& + \Big( \frac{5\alpha^3}{72m_g^{10}} {\bf J}_1 - \frac{97\alpha^3-416\alpha^2\gamma}{648m_g^{10}} {\bf J}_2 + \frac{53\alpha^3-496\alpha^2\gamma+2304\alpha\gamma^2-6144\gamma^3}{216m_g^{10}}{\bf J}_3 \Big) l^{10} \nn\\ 
\fl&& + \Big( \frac{\alpha^3}{162m_g^{12}} {\bf J}_1 - \frac{3\alpha^3-16\alpha^2\gamma}{162m_g^{12}} {\bf J}_2 + \frac{\alpha^3+48\alpha^2\gamma-576\alpha\gamma^2+1536\gamma^3}{162m_g^{12}}{\bf J}_3 \Big) l^{12} \Big\} \nn\\
\fl&&+ {\rm terms\; with\;} p_1 {\rm \;and\;} p_2 \nn\\
\fl&=& \frac{\alpha^2}{\epsilon}\frac{5i\lambda^3m_g^4}{442368\pi^2M_P^3}\Big\{ 13\alpha{\bf J}_1 - \frac{1}{3}(35\alpha+32\gamma){\bf J}_2 + (11\alpha+16\gamma){\bf J}_3 \Big\} \nn\\
\fl&&+ {\rm terms\; with\;} p_1 {\rm \;and\;} p_2 + {\rm finite}\,.
\eea
Note that in the second equality of (\ref{eqn:3ptftnint}), instead of $(l^2 + m_g^2)^{-3}$ there should have been $\{l^2 + m_g^2 + x(1-x)p_1^2+y(1-y)p_2^2-2xyp_1\cdot p_2\}^{-3}$ with the appropriate integrations over the Feynman parameters $x$ and $y$. But here we are only interested in the contribution in the form of the tree level 3-point vertex, and take the shortcut of ignoring any dependence on the external momenta $p_1$ and $p_2$.

\section{Graviton loops with a more general interaction}\label{ggloops}
An interaction more general than (\ref{eqn:cubicint}) may contain derivatives, and a reasonable way to introduce such interactions is to expand $\sqrt{-g}R$ to higher orders in $h$. That is, we now investigate
\bea\label{eqn:4thopfa}
\fl S_{m_g} &=& S_{\rm PF} 
+ 2M_P^2\int\d^4x \,\Big\{ [ \sqrt{-g}\,R\,]\big|_{h^3} + [ \sqrt{-g}\,R\,]\big|_{h^4} \nn\\
\fl&&- \frac{m_g^2}{4(2!)^33!M_P^3}\big( x_1 h^\mu_\nu h^\nu_\sigma h^\sigma_\mu + x_2 h_{\mu\nu}h^{\mu\nu}h
+ x_3 h^3 \big) \nn\\
\fl&&- \frac{m_g^2}{4(2!)^44!M_P^4}\big( y_1 h^\mu_\nu h^\nu_\sigma h^\sigma_\lambda h^\lambda_\mu 
+ y_2 (h_{\mu\nu}h^{\mu\nu})^2 + y_3 h^\mu_\nu h^\nu_\sigma h^\sigma_\mu h
+ y_4 h_{\mu\nu}h^{\mu\nu} h^2 + y_5 h^4 \big) \Big\} \nn\\
\fl&=& S_{\rm PF} + \int\d^4x \,\Big( \frac{h^{\mu_1\nu_1} h^{\mu_2\nu_2} h^{\mu_3\nu_3}}{M_P} 
V_g^{(3)}{}_{\mu_1\nu_1;\mu_2\nu_2;\mu_3\nu_3} \nn\\
\fl&&\hspace{50pt}+ \frac{h^{\mu_1\nu_1} h^{\mu_2\nu_2} h^{\mu_3\nu_3} h^{\mu_4\nu_4}}{M_P^2} V_g^{(4)}{}_{\mu_1\nu_1;\mu_2\nu_2;\mu_3\nu_3;\mu_4\nu_4} \Big) \,.
\eea
Now that $M_P^{-1}$ plays the role of the coupling, we need the cubic and quartic vertices in order to get loops of ${\mathcal O}(M_P^{-2})$. Explicit form of $\sqrt{-g}\,R\big|_{h^3}$ and $\sqrt{-g}\,R\big|_{h^4}$ can be found at \eg, \cite{gvertices}, and then we obtain  $V_g^{(3)}$ and $V_g^{(4)}$ straightforwardly.

To find loop corrections to $S_{\rm PF}$, we path-integrate over $h$, while ${\cal O}(h^3)$ and ${\cal O}(h^4)$ terms provide quantum interactions:
\bea\label{eqn:pfpf}
\fl Z[J] &=& \int {\cal D}h\, \exp\Big[\,i \Big(S_{m_g} 
+ \int \d^4 x\, J^{\mu\nu}h_{\mu\nu} \Big) \Big] \nn\\
\fl &=& {\cal N} \exp\Big[\, i \int\d^4 z \Big( \frac{iV_g^{(3)}}{M_P}\frac{\delta^3}{\delta J(z)^3}
+ \frac{V_g^{(4)}}{M_P^2}\frac{\delta^4}{\delta J(z)^4} \Big) \Big] \nn\\
\fl&&\qquad\times\exp\Big[\, \frac{i}{2}\int \d^4 x \d^4 y J(x) \big(-i\tilde{\bf P}_{m_g}^{(0)}(x-y)\big) J(y) \Big]\,,
\eea
where ${\cal N}$ is a normalization constant and $\tilde{\bf P}_{m_g}^{(0)}$ 
is the inverse Fourier transform of the tree level PF graviton propagator, (\ref{eqn:tlpfgp}). With 
\be
\tilde{\bf P}_{m_g}^{(0)} = \includegraphics[width=1cm]{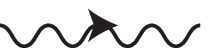} \,,
\quad
i\int\d^4x J(x) \big(-i\tilde{\bf P}_{m_g}^{(0)}(x-y)\big) = 
\includegraphics[width=1cm]{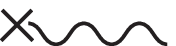} \,,
\ee
we can expand (\ref{eqn:pfpf}) diagrammatically: Up to ${\mathcal O}(M_P^{-3})$, 
\vspace{5pt}
\bea\label{eqn:lmgi}
\fl 3i \includegraphics[bb=-5 15 70 50,keepaspectratio=true,width=1.6cm]{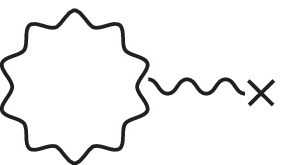}
\quad + 6 \includegraphics[bb=-5 15 70 50,keepaspectratio=true,width=1.6cm]{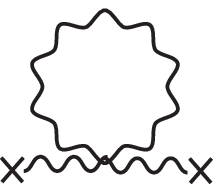}
- \frac{1}{2} \Big( 18 \includegraphics[bb=-5 15 70 50,keepaspectratio=true,width=1.6cm]{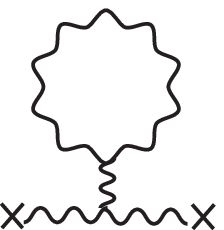} 
+ 18 \includegraphics[bb=-5 15 70 50,keepaspectratio=true,width=1.6cm]{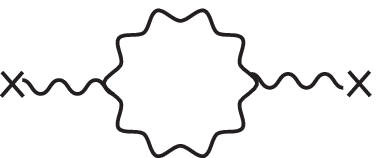} \qquad\; \Big) \,, \\
\fl \hspace{15pt}\underbrace{\hspace{20pt}}_{(a)} \hspace{56pt}\underbrace{\hspace{20pt}}_{(b)} \hspace{60pt}\underbrace{\hspace{20pt}}_{(c)} \hspace{59pt}\underbrace{\hspace{20pt}}_{(d)} \nn
\eea
where $(a)\sim(d)$ are the loop parts of corresponding diagrams. The linear tadpole, $(a)$, is
\bea
(a) &=& \frac{5i m_g^4}{768\pi^2 M_P\epsilon} (-2+3x_1+4x_2)\eta_{\alpha\beta} + {\rm finite} \,.
\eea
In order not to have a tadpole, we have to fix $x_2=\frac{2-3x_1}{4}$, which in turn makes $(c)$ vanish. The rest of the loops are
\bea\label{eqn:(b)}
\fl (b) &=& \frac{5i m_g^2}{13824\pi^2 M_P^2\epsilon} \Big\{\Big( 18k^2+(4y_1-4y_2+27y_3+36y_4)m_g^2\Big){\bf I}_1 \nn\\
\fl &&\hspace{60pt}-\Big(9k^2-(9+19y_1+44y_2)m_g^2\Big){\bf I}_2 + 9{\bf I}_3 - 18{\bf I}_4 \Big\} + {\rm finite} \,, \\
\label{eqn:(d)}
\fl (d) &=& \frac{5im_g^2}{1990656\pi^2M_P^2\epsilon} \Big\{ \Big((600+312x_1+63x_1^2)k^2-(376-348x_1+9x_1^2)m_g^2\Big){\bf I}_1 \nn\\
\fl&&- \Big((272+456x_1+45x_1^2)k^2-2(376-348x_1+9x_1^2)m_g^2\Big){\bf I}_2 \nn\\
\fl&&+ 18(28+4x_1+9x_1^2){\bf I}_3 - 2(428+102x_1+81x_1^2){\bf I}_4 \Big\} + {\mathcal O}\Big(\frac{k^4}{M_P^2}\Big) + {\rm finite} \,.
\eea
Applying $\delta^2 / (i\delta J^{\mu\nu})(i\delta J^{\lambda\rho})$ to (\ref{eqn:lmgi}) gives the 1-loop correction to the 2-point 1PI:
\bea\label{eqn:mgi2p1pi}
\fl&&12(b) - 18(d) = \frac{5im_g^2}{110592\pi^2M_P^2\epsilon} \Big[ \Big\{3(376-104x_1-21x_1^2)k^2 \nn\\
\fl&&\hspace{80pt}-\big(348x_1-9x_1^2-8(47+48y_1-48y_2+324y_3+432y_4)\big)m_g^2\Big\}{\bf I}_1 \nn\\
\fl&&\hspace{20pt}- \Big\{(592-456x_1-45x_1^2)k^2-2(56+348x_1-9x_1^2+912y_1+2112y_2)m_g^2\Big\}{\bf I}_2 \nn\\
\fl&&\hspace{20pt}+ 18(20-4x_1-9x_1^2){\bf I}_3 - 2(436-102x_1-81x_1^2){\bf I}_4 \Big] + {\mathcal O}\Big(\frac{k^4}{M_P^2}\Big) + {\rm finite} \,.
\eea
Then,
\bea
\fl&&\hat h^{\mu\nu} \big\{ 12(b) - 18(d) \big\} \hat h^{\lambda\rho} \nn\\
\fl&&\propto \big\{9(k^0)^2(40-81x_1+33x_1^2) - (k^i)^2(56 - 600x_1 - 27x_1^2) + \cdots \big\} h_{00}{}^2 + \cdots \,,
\eea
which does not vanish for any $x_1$.

\section{Summing up 1PI's}\label{resum}
First of all, we need to know various contractions between ${\bf I}_i$'s, which are worked out 
in table \ref{table1}. Here tensors on the leftmost column with $(\alpha\beta;\mu\nu)$ index structure 
are multiplied into those on the top row with $(\mu\nu;\lambda\rho)$, and 
${\bf I}_{41} = \eta_{\alpha\beta}k_\mu k_\nu$ or $\eta_{\mu\nu}k_\lambda k_\rho$, 
${\bf I}_{42}=k_\alpha k_\beta \eta_{\mu\nu}$ or $k_\mu k_\nu \eta_{\lambda\rho}$.  
\begin{table}[ht]
\caption{Tensor multiplication table}
\begin{center}
\begin{tabular}{c|c c c c c c}
      & $\quad {\bf I}_1\quad$ & $\quad {\bf I}_2\quad$ & $\quad {\bf I}_3\quad$ & $\quad {\bf I}_{41}\quad$ 
      & $\quad {\bf I}_{42}\quad$ & $\quad {\bf I}_5\quad$ \\ 
\hline                  
${\bf I}_1$ & $4{\bf I}_1$ & $2{\bf I}_1$ & $4{\bf I}_{41}$ & $4{\bf I}_{41}$ & $k^2 {\bf I}_1$ & $k^2 {\bf I}_{41}$ \\ 
${\bf I}_2$ & $2{\bf I}_1$ & $2{\bf I}_2$ & $2{\bf I}_3$ & $2{\bf I}_{41}$ & $2{\bf I}_{42}$ & $2{\bf I}_5$ \\
${\bf I}_3$ & $4{\bf I}_{42}$ & $2{\bf I}_3$ & $2k^2{\bf I}_3+8{\bf I}_5$ & $4{\bf I}_5$ & $4k^2{\bf I}_{42}$ & $4k^2{\bf I}_5$ \\
${\bf I}_{41}$ & $k^2{\bf I}_1$ & $2{\bf I}_{41}$ & $4k^2{\bf I}_{41}$ & $k^2{\bf I}_{41}$ & $k^4{\bf I}_1$ & $k^4{\bf I}_{41}$ \\
${\bf I}_{42}$ & $4{\bf I}_{42}$ & $2{\bf I}_{42}$ & $4{\bf I}_5$ & $4{\bf I}_5$ & $k^2{\bf I}_{42}$ & $k^2{\bf I}_5$ \\
${\bf I}_5$ & $k^2{\bf I}_{42}$ & $2{\bf I}_5$ & $4k^2{\bf I}_5$ & $k^2{\bf I}_5$ & $k^4{\bf I}_{42}$ & $k^4{\bf I}_5$ 
\label{table1}
\end{tabular}
\end{center}
\end{table}
A graviton self energy, ${\bf \Pi}_{\alpha\beta;\lambda\rho}$, has the general form of
\be\label{eqn:gse}
{\bf \Pi} = \sum_{j=1}^5 b_j(k) {\bf I}_j\,,
\ee
because ${\bf I}$'s are a complete basis.

\subsection{GR}
Let us look at the case of GR first. To do the summation 
\be\label{eqn:grrsprops}
{\bf P} = {\bf P}^{(0)} + {\bf P}^{(0)} \cdot {\bf\Pi} \cdot {\bf P}^{(0)} 
+ {\bf P}^{(0)} \cdot ({\bf\Pi} \cdot {\bf P}^{(0)})^2 + \cdots \,,
\ee
with the graviton propagator in the de Donger gauge,
\be\label{eqn:treegrprop}
{\bf P}^{(0)} = \frac{i}{k^2}\Big( -\frac{{\bf I}_1}{2} + \frac{{\bf I}_2}{2} \Big)\,,
\ee 
we should find the $n$-dependence of $({\bf\Pi} \cdot {\bf P}^{(0)})^n$. Using table \ref{table1}, we get
\be
{\bf\Pi} \cdot {\bf P}^{(0)} = \sum_i r_i {\bf I}_i\,,
\ee 
with
\bea\label{eqn:mf}
&r_1 = \frac{b_1+b_2}{k^2}+ \frac{b_4}{2}\,,\quad
r_2 = -\frac{b_2}{k^2}\,,\quad
r_3 = -\frac{b_3}{k^2}\,,\quad \nn\\
&r_{41} = -\frac{b_4}{k^2}\,, \quad
r_{42} = \frac{2b_3+b_4}{k^2}+\frac{b_5}{2}\,,\quad
r_5 = -\frac{b_5}{k^2}\,.
\eea
Then, by writing
\be
({\bf\Pi} \cdot {\bf P}^{(0)})^{n+1} = \sum_i a_i^{(n)} {\bf I}_i
= \sum_i r_i {\bf I}_i \cdot \sum_j a_j^{(n-1)} {\bf I}_j \,,
\ee
we obtain the recurrence relations between $a_i^{(n)}$ and $a_j^{(n+1)}$:
\bea
\label{eqn:a}
\fl a_1^{(n+1)} &=& (4r_1+2r_2+r_{41}k^2)a_1^{(n)} + 2r_1 a_2^{(n)}+ (r_1+r_{41}k^2)k^2 a_{42}^{(n)} \,,\\
\label{eqn:b}
\fl a_2^{(n+1)} &=& 2r_2 a_2^{(n)}\,,\\
\label{eqn:c}
\fl a_3^{(n+1)} &=& 2r_3 a_2^{(n)} + 2(r_2+r_3k^2)a_3^{(n)}\,,\\
\label{eqn:d}
\fl a_{41}^{(n+1)} &=& 2r_{41}a_2^{(n)} + 4(r_1+r_{41}k^2)a_3^{(n)} \nn\\
\fl &&\quad+(4r_1+2r_2+r_{41}k^2)a_{41}^{(n)} + (r_1+r_{41}k^2)k^2a_5^{(n)}\,,\\
\label{eqn:e}
\fl a_{42}^{(n+1)} &=& (4r_3+4r_{42}+r_5k^2)a_1^{(n)}+ 2r_{42}a_2^{(n)} \nn\\
\fl &&\quad+(2r_2+(4r_3+r_{42}+r_5k^2)k^2)a_{42}^{(n)} \,,\\
\label{eqn:f}
\fl a_5^{(n+1)} &=& 2r_5a_2^{(n)} + 4(2r_3+r_{42}+r_5k^2)a_3^{(n)} \nn\\
\fl &&+(4r_3+4r_{42}+r_5k^2)a_{41}^{(n)} +(2r_2+(4r_3+r_{42}+r_5k^2)k^2)a_5^{(n)}\,.
\eea
(\ref{eqn:b}) and (\ref{eqn:c}) are trivial to solve. With $a_2^{(n)}$ and $a_3^{(n)}$ determined, the remaining equations are grouped into two sets of coupled equations: (\ref{eqn:a}) and (\ref{eqn:e}), and (\ref{eqn:d}) and (\ref{eqn:f}). Each of these sets can be solved in the same way as solving the Fibonacci sequence. The complete answer is rather lengthy:
\bea
\fl a_1^{(n)} &=& \frac{\gamma_+ Y_-^{(n)}-\gamma_- Y_+^{(n)}}{\gamma_+-\gamma_-} \,,\quad
a_2^{(n)} = \frac{1}{2} (2r_2)^{n+1} \,,\quad
a_3^{(n)} = \frac{(2r_2+2r_3k^2)^{n+1}-(2r_2)^{n+1}}{2k^2} \,,\nn\\
\fl a_{41}^{(n)} &=& \frac{\gamma_+ X_-^{(n)}-\gamma_- X_+^{(n)}}{\gamma_+-\gamma_-} \,,\quad
a_{42}^{(n)} = \frac{Y_+^{(n)}-Y_-^{(n)}}{\gamma_+-\gamma_-}  \,,\quad
a_5^{(n)} = \frac{X_+^{(n)}-X_-^{(n)}}{\gamma_+-\gamma_-} \,,
\eea
where
\bea\label{eqn:gamma}
\fl\gamma_\pm &=& \frac{-\alpha_1+\alpha_2'\pm\sqrt{(\alpha_1-\alpha_2')^2+4\alpha_1'\alpha_2}}{2\alpha_1'}\,,\\
\fl X_\pm^{(n)} &=& (r_{41} + \gamma_\pm r_5)\sigma_\pm^n +\rho_{1\pm}\frac{\sigma_\pm^n-(2r_2)^n}{\sigma_\pm-2r_2}
+\rho_{2\pm}\frac{\sigma_\pm^n-(2r_2+2r_3k^2)^n}{\sigma_\pm-(2r_2+2r_3k^2)}\,,\\
\fl Y_\pm^{(n)} &=& (r_1 + \gamma_\pm r_{42})\sigma_\pm^n +\rho_{\pm}\frac{\sigma_\pm^n-(2r_2)^n}{\sigma_\pm-2r_2}\,,
\eea
with
\bea\label{eqn:therest}
\fl &&\alpha_1 = 4r_1+2r_2+r_{41}k^2 \,,\quad \alpha_2 = (r_1+r_{41}k^2)k^2\,, \quad\nn\\
\fl &&\alpha_1' = 4r_3+4r_{42}+r_5k^2 \,,\quad \alpha_2' = 2r_2+(4r_3+r_{42}+r_5k^2)k^2 \,, \nn\\
\fl &&\sigma_\pm = \frac{\alpha_1+\alpha_2'\pm\sqrt{(\alpha_1-\alpha_2')^2+4\alpha_1'\alpha_2}}{2}\,,\quad
\rho_{\pm} = 2r_2(r_1+\gamma_\pm r_{42}) \,,\\
\fl &&\rho_{1\pm} = -2r_2\Big\{r_{41}+\gamma_\pm r_5+\frac{2}{k^2}(r_1+\gamma_\pm(2r_3+r_{42})) \Big\} \,, \quad\nn\\
\fl &&\rho_{2\pm} = 4\Big(\frac{r_2}{k^2}+r_3\Big)\Big\{r_1+r_{41}k^2+\gamma_\pm(2r_3+r_{42}+r_5k^2) \Big\} \,. \nn
\eea
Then,
\bea
\fl &&{\bf P}^{(0)} \cdot ({\bf\Pi} \cdot {\bf P}^{(0)})^{n+1} \nn\\
\fl &&= \frac{i}{2(\gamma_+-\gamma_-)k^2}
\Big[\; \frac{\rho_{+}(2^nr_2^n-\sigma_+^n)(2\gamma_--k^2)}{2r_2-\sigma_+} 
- \frac{\rho_{-}(2^nr_2^n-\sigma_-^n)(2\gamma_+-k^2)}{2r_2-\sigma_-} \nn\\
\fl&&\quad- (2r_2)^{n+1} (\gamma_+-\gamma_-) + \gamma_+^n(r_1+r_{42}\gamma_+)(2\gamma_--k^2)
- \gamma_-^n(r_1+r_{42}\gamma_-)(2\gamma_+-k^2) \,\Big] {\bf I}_1 \nn\\
\fl &&\quad+ \frac{1}{2k^2}(2r_2)^{n+1}{\bf I}_2 + ({\bf I}_3\,,\;{\bf I}_4\,,\;{\bf I}_5) \,.
\eea
As this is merely a geometric series, summing them up is straightforward. 
Plugging (\ref{eqn:gamma}) and (\ref{eqn:therest}), we finally get
\bea\label{eqn:grloopsum}
\fl {\bf P} &=& {\bf P}^{(0)} + \sum_{n=0}^\infty{\bf P}^{(0)} \cdot ({\bf\Pi} \cdot {\bf P}^{(0)})^{n+1} \nn\\
\fl&=& - \frac{i}{k^2+2b_2} \frac{{\bf I}_1}{3} + \frac{i}{k^2+2b_2} \frac{{\bf I}_2}{2} 
- i\frac{2b_5k^4+(1+8b_3+4b_4)k^2+2(b_1+2b_2)}{d(k)} \frac{{\bf I}_1}{3} \nn\\
\fl&&+ ({\bf I}_3\,,\;{\bf I}_4\,,\;{\bf I}_5) \,,
\eea
with
\bea
d(k) &=& b_5k^6+\big\{2+4b_3-2b_4(2-3b_4)-2b_5(3b_1+2b_2)\big\}k^4 \nn\\
&&-8\big\{b_1(1+3b_3)+b_2(2b_3+b_4)\big\}k^2-8b_2(2b_1+b_2)\,.
\eea
Since GR has a full GCI, ${\bf \Pi}$ should be invariant under $h_{\mu\nu} \to h'_{\mu\nu} = h_{\mu\nu} + \partial_\mu\xi_\nu + \partial_\nu\xi_\mu$, \ie, 
\bea
\fl &&0 = \int \d^4 x\, \big\{{\bf \Pi}(h'_{\mu\nu}) - {\bf \Pi}(h_{\mu\nu}) \big\}
= -2\int\d^4 x\,\xi_\mu\partial_\nu \frac{\delta}{\delta h_{\mu\nu}} {\bf \Pi} \nn\\
\fl &&\Rightarrow\;0 = (b_1+k^2b_4)k_\nu \hat h + 2(b_2 + k^2b_3)k_\rho \hat h^\rho_\nu + (2b_3 + b_4 + k^2 b_5) k_\nu k_\lambda h_\rho \hat h^{\lambda\rho}\,,
\eea
which gives
\be
b_1+k^2b_4=0\,,\quad b_2 + k^2b_3=0\,,\quad 2b_3 + b_4 + k^2 b_5=0\,,
\ee
and (\ref{eqn:grloopsum}) turns into
\bea\label{eqn:grloopsumwwi}
\fl {\bf P} = \frac{i}{k^2+2b_2} \Big\{\frac{-k^2+b_1}{2k^2-3b_1-2b_2}{\bf I}_1 + \frac{{\bf I}_2}{2} + \frac{b_2}{k^4}{\bf I}_3 - \frac{b_1+2b_2}{k^2(2k^2-3b_1-2b_2)}\Big({\bf I}_4 + \frac{2}{k^2}{\bf I}_5\Big)\Big\} \,. \nn\\
\eea
In GR, the forms of the quadratic action and the corresponding propagator can vary as we change the gauge choice, so that a difference between the resummed propagator and the tree level one may not be a problem as long as there is a general coordinate transformation that connects them.

\subsection{PF}
For PF theory, we can follow the same steps as those of the GR case. 
The differences are that there is no Ward identity here and that we use ${\bf P}_{m_g}^{(0)}$ instead of ${\bf P}^{(0)}$, which changes (\ref{eqn:mf}) into
\bea
\fl r_1 &=& \frac{b_1}{3m_g^2}+ \frac{2b_2}{3(k^2+m_g^2)}+\frac{b_4k^2}{3m_g^2}\,,\quad
r_2 = -\frac{b_2}{k^2+m_g^2}\,,\quad
r_3 = -\frac{b_2}{m_g^2(k^2+m_g^2)}-\frac{b_3}{m_g^2}\,,\nn\\
\fl r_{41} &=& -\frac{2b_1}{3m_g^4}+\frac{2b_2}{3m_g^2(k^2+m_g^2)}-\frac{b_4(2k^2+3m_g^2)}{3m_g^4}\,, \quad\nn\\
\fl r_{42} &=& \frac{2b_2}{3m_g^2(k^2+m_g^2)}+\frac{4b_3}{3m_g^2}+\frac{b_4}{3m_g^2}+\frac{b_5k^2}{3m_g^2}\,, \\
\fl r_5 &=& -\frac{4b_2}{3m_g^4(k^2+m_g^2)}-\frac{8b_3}{3m_g^4}-\frac{2b_4}{3m_g^4}-\frac{b_5(2k^2+3m_g^2)}{3m_g^4}\,. \nn
\eea
Then, the resummed PF graviton propagator is
\bea
\fl {\bf P}_{m_g} &=& {\bf P}_{m_g}^{(0)} 
+ \sum_{n=0}^\infty{\bf P}_{m_g}^{(0)} \cdot ({\bf\Pi} \cdot {\bf P}_{m_g}^{(0)})^{n+1} \nn\\
\fl &=& -i \Big\{\frac{1}{k^2+m_g^2+2b_2} + \frac{b_5k^4+2(2b_3+b_4)k^2+b_1+2b_2}{d_{m_g}\hspace{-2pt}(k)}\Big\} \frac{{\bf I}_1}{3} + \frac{i}{k^2+m_g^2+2b_2} \frac{{\bf I}_2}{2} \nn\\
\fl &&- \frac{i}{2k^2}\Big(\frac{1}{k^2+m_g^2+2b_2} - \frac{1}{2b_3k^2+m_g^2+2b_2}\Big){\bf I}_3 
+ \frac{i n_{m_g}\hspace{-2pt}(k)}{(k^2+m_g^2+2b_2)d_{m_g}\hspace{-2pt}(k)}{\bf I}_4 \nn\\
\fl &&+ \frac{i}{3k^4}\Big\{ \frac{2}{k^2+m_g^2+2b_2} - \frac{6}{2b_3k^2+m_g^2+2b_2} \nn\\
\fl &&\hspace{40pt}- \frac{b_5k^4-(6-4b_3-8b_4)k^2+16b_1+8b_2-12m_g^2}{d_{m_g}\hspace{-2pt}(k)} \Big\}{\bf I}_5 \,,
\eea
where
\bea
d_{m_g}\hspace{-2pt}(k) &=& 2b_5k^6+\big(8b_3+b_4(4+3b_4)-b_5(3b_1+2b_2)+2b_5 m_g^2\big)k^4 \nn\\
&&+\big(2b_1(1-6b_3)+4b_2(1-2b_3-b_4)+2(4b_3-b_4)m_g^2\big)k^2\nn\\
&&-(4b_1+2b_2-3m_g^2)(2b_2+m_g^2) \,, \\
n_{m_g}\hspace{-2pt}(k) &=& b_5 k^4 + (4b_3+3b_4+b_4^2-b_1b_5+b_5m_g^2)k^2 \nn\\
&&+ b_1(2-4b_3) + 2b_2(1+b_4)-(1-4b_3-b_4)m_g^2 \,.
\eea

\end{document}